\UseRawInputEncoding
\documentclass[UTF8,twocolumn,preprintnumbers,amsmath,amssymb]{revtex4}

\usepackage{graphicx}
\usepackage{dcolumn}
\usepackage{bm}
\begin{document}

\preprint{}

\title{The anomalous process in singlet fission kinetic model with time-dependent coefficient}

\author{Fang-Qi Hu}
\thanks{Email: thufang2008@163.com}
\affiliation{College of Physics and Electronic Engineering, Northwest Normal University, Lanzhou 730070, China.}
\author{Zi-Fa Yu}
\affiliation{College of Physics and Electronic Engineering, Northwest Normal University, Lanzhou 730070, China.}
\author{Ji-Ming Gao}
\affiliation{College of Physics and Electronic Engineering, Northwest Normal University, Lanzhou 730070, China.}
\author{Ju-Kui Xue}
\thanks{Email: xuejk@163.com}
\affiliation{College of Physics and Electronic Engineering, Northwest Normal University, Lanzhou 730070, China.}

\begin{abstract}
In the third generation photovoltaic device, the main physical mechanism that the photoelectric conversion efficiency is enhanced is the singlet fission (SF). In order to accurately describe the SF and reveal physical process in theoretically, we introduce the anomalous process (AP) in SF dynamics based on previous model, including anomalous fission, anomalous fusion, anomalous dissociation and combination of triplet pair states, anomalous decay and diffusion of single triplet exciton. The effects of the AP on SF are investigated by the kinetic model with time-dependent coefficient. Further, according to the results we make the optimal simulations for the experimental data [G. B. Piland et al., J. Phys. Chem. C, 2013, 117, 1224] by adjusting the rate coefficients and exponents in the mended kinetic equations. The results show that the model considered AP is more accurate than previous that to describe SF dynamics, demonstrating that the AP do exist in SF. The model also provides the theoretical foundation for how varies experimentally physical factors to make SF occur to tend to required direction.
\end{abstract}


\maketitle


Harvesting solar energy through photocells is one of the potential approaches that solve world energy requirements. In the past decade, the studies of some organic materials with singlet fission (SF), including $\pi$-conjugated molecular crystals and polymers\cite{G.B.Piland.117.2013,ma2013fluorescence,beljonne2013charge,musser2015evidence,berkelbach2013microscopic,berkelbach2014microscopic,wilson2013temperature},
were brought to a new climax due to its character with high quantum yield of triplet exciton in photoelectric conversion. SF is a process that can lead to multiple exciton generation by absorbing one photon in some organic material \cite{singh1963double,S.Singh.42.1965,A.Rao.132.2010,C.Wang.132.2010,C.Ramanan.134.2012}. Therefore, the photoelectric conversion efficiency of the photovoltaic devices that is made of the organic material with SF can surpass the Schockley--Queisser efficiency limit of traditional solar cell, which was largely proven \cite{shockley1961detailed,hanna2006solar,congreve2013external,thompson2013slow,pazos2017silicon}.
It motivated intense research in theory and experiment, in which the researchers devoted to reveal the mechanism of and deepen our understanding of SF \cite{hochstrasser1962luminescence,kepler1963triplet,yago2016magnetic,groff1970coexistence,merrifield1968theory,greyson2010maximizing,zimmerman2011mechanism,irkhin2011direct,G.B.Piland.117.2013,J.J.Burdett.585.2013,J.J.Burdett.135.2011},
to raise the photoelectric conversion efficiency and the stability of the material \cite{chen2014effects,fallon2019exploiting,conrad2019controlling},
and to design novel SF material and devices\cite{ehrler2012singlet,jadhav2012triplet,L.Yang.15.2015,paci2006singlet,bendikov2004oligoacenes,ryerson2014two}.

The magnetic field effect (MFE) is powerful tool to identify the SF material, in which the fluorescence decay (FD) dynamics in SF depend on the strength and orientation of magnetic field\cite{G.B.Piland.117.2013,J.J.Burdett.585.2013,yago2016magnetic}. In fact, the application of an external magnetic field effects the yields of triplet pair from fission and the dynamics of prompt and delayed fluorescence. The effects can be described by the quantum mechanical theory developed by Johnson and Merrifield \cite{merrifield1968theory,johnson1970effects},
and are investigated by extended model in later\cite{hu2019improved,G.B.Piland.117.2013,J.J.Burdett.585.2013}. According the theory, its source is that magnetic field produces the Zeeman splitting and varies zero-field splitting in spin Hamiltonian, mixing the spin eigenstates of exciton pair system. More and more experimental evidences shown that the theory accounted well for the MFE of SF.
However, a variety of improved and extended model were developed and work west based on the theory due to its deficiency, such as the introduction of the separate and combination of triplet pair, the hopping of single triplet exciton. In spite of these meticulous consideration for SF process, a perfect fitting of the fluorescence dynamics is difficult, and the rate constant obtained by the theory is not consistent with that measured experimentally in the order of magnitudes\cite{hu2019improved,dillon2013different}.

Recently, Yao Yao come up with a behaviour analogous to the subdiffusion process in order to explain a smaller-than-unity exponential decay of spin state of exciton pair in SF simulated by full-quantum method \cite{yao2016coherent}. The process was described by a dynamic differential equation, in which time dependent dynamics coefficient was introduced as $k_{\textrm{SF}}\left(\sim t^{\nu-1}\right)$, and has been observed in experiment\cite{zhang2014nonlinear}. Commonly, the diffusion refers to the mobilities of protein in cell\cite{klafter2005anomalous}, the motion of Brownian particles in solution \cite{mandelbrot1968fractional},
the random walk of excitons in crystals\cite{burdett2010excited,tamai2015exciton}
etc.. Nevertheless, the subdiffusion is an anomalous diffusion, to which the diffusion transition in the presence of the exciton traps in crystal\cite{akselrod2014visualization,sha2019dynamical}. In this paper we will call it as one of the anomalous process (AP) because of the difference between the diffusion and the SF, and the other processes, or between the subdiffusion and the smaller-than-unity decay exponential behaviour in SF. Surely, it is necessary to introduce the APs in whole SF process based on the expanded model\cite{hu2019improved}, including anomalous fission and decay of overall singlet states, anomalous dissociation and fusion of triplet pair states, anomalous combination and diffusion of single triplet excitons. We analyse the effects of each AP on fluorescence decay dynamics by the new kinetic model. The results show that the effects of each parameter corresponding to different APs on SF are distinct. According to it, the more perfect fitting to previous experiment result completed by Piland et al. is presented using the model with these APs than without. The results fitted manifest directly that our model amended is more precise than previous, and indirectly that the APs in SF exist. 

This paper is organized as follows. Sec. \ref{sec2} introduces theory and model on the system studied and consists of two parts: In SubSec. \ref{subsec2.1}, the diffusion theory of exciton in crystal is simply discussed. In SubSec. \ref{subsec2.2}, the spin Hamiltonian for the system is illustrated. In SubSec. \ref{subsec2.3}, a kinetic model with time dependent dynamics coefficient is introduced based on previous extended model on SF. In Sec. \ref{sec3}, we give the results. In SubSec. \ref{subsec3.1}, the effects of each AP introduced on SF dynamics are investigated in detail for total random molecular system in the presence of an external magnetic field. In SubSec. \ref{subsec3.2}, the optimal fittings of fluorescence decay to previous experimental results are shown by the kinetic model amended. Sec. \ref{sec4} is a brief summary and future perspective.

\section{Theory and Model}\label{sec2}

\subsection{the diffusion}\label{subsec2.1}
To fit the SF dynamics obtained by full-quantum dynamical simulation, Yao Yao put forward a behaviour analogous to the subdiffusion with a smaller-than-unity exponential decay in simulation\cite{yao2016coherent}. Therefore, the time dependent dynamics coefficient was introduced in a density matrix equation governing singlet. In fact, the behaviour of evolution of singlet is similar to the subdiffusion, but different. In order to introduce the AP in our model, here we must state the related knowledge and theory of the diffusion and subdiffusion of exciton.

As we known, the pathway of solar energy harvesting is that the excitation energy is transported to interfaces in assemblies of functional organic molecules in organic solar cells \cite{zhang2014molecular,pensack2016observation}. 
However, in the SF organic material the carriers transporting energy are the uncorrelated triplet excitons generated by correlated triplet pairs separating\cite{kohler2009triplet}. Therefore, the exciton transport is the core of photoelectric conversion in nanostructured thin films and crystals \cite{escalante2010long}, which governs the efficiency of nanostructured optoelectronic devices,  including molecular, polymeric and colloidal quantum dot solar cells \cite{lunt2009exciton,menke2013tailored}, light-emitting diodes \cite{hofmann2012singlet} and excitonic transistors \cite{high2008control}.
Using tetracene as an archetype molecular crystal, the studies shown that the transit mode in crystals is accomplished by the transit of the localized excitation to a neighbouring molecular site, and is described by a hopping process with random walk of exciton between neighbouring molecular sites due to F�rster-like or Dexter energy transfer mechanisms \cite{soos1972generalized,akselrod2014visualization}. 
The exciton diffusion tend to the lowest energy sites and thus is largely pre-defined by the energetic landscape \cite{herz2004time}.
Therefore, the exciton diffusion is a key process that can affect the energy conversion efficiency of organic solar cell.

The exciton diffusion process in isotropy medium is described by the well-known diffusion equation in any dimensionless $x$ \cite{crank1979mathematics},
\begin{equation}
\frac{\partial n\left(x,t\right)}{\partial t}=D\frac{\partial^{2}n\left(x,t\right)}{\partial x^{2}}-\frac{n\left(x,t\right)}{\tau}-\frac{\gamma n^{2}\left(x,t\right)}{\sqrt{\tau}},\label{eq:diffusion}
\end{equation}
where $n\left(x,t\right)$ is the probability density distribution at a time $t$ and position $x$, $D$ is the time dependent diffusivity of exciton in the $x$ direction, $\tau$ is the natural decay lifetime of the exciton, and the $\gamma$ is a coefficient that accounts for exciton-exciton annihilation. Eq.(\ref{eq:diffusion}) describes the random motion of the neutral exciton that spread from the sites of high concentration to the sites of low concentration.
Here we only concern the form of the equation, but not the solution. For convenience, the initial time is set as $t_{0}=0$, the diffusivity $D$ is defined as \cite{pandya2021exciton}
\begin{equation}
D(t)=\dfrac{A}{2}t^{\alpha-1},\label{eq:diffusivity}
\end{equation}
where $A$ is scaling factor, and $\alpha$ is diffusion exponent which accounts for the nonlinearity in time. $A$ and $\alpha$ are key parameters for device optimization \cite{menke2014exciton,mikhnenko2015exciton}, and can be empirically extracted by fitting the equation to best match the experimental data \cite{sung2020long}. The two parameters are related to the band structure \cite{silins1994organic}, the dispersive transport of exciton \cite{blom1998dispersive}, the local part of the interaction, packing way, symmetry of exciton \cite{pandya2021exciton}, and the power of excitation light \cite{wittmann2020energy}. Because exciton-exciton annihilation can accelerate the exciton dynamics, it can be employed to measure diffusion coefficients in experiment \cite{jang2021excitons}. Besides, the time-resolved microscopy was used to probe the exciton transport \cite{zhu2016two}.

In Eq.(\ref{eq:diffusivity}) the value of $\alpha$ has three conditions according to the behaviour of exciton transport. For $\alpha=1$, the exciton transport behaves the normal diffusion, in which the diffusivity $D$ is time independent. However, $D$ becomes time dependent for $\alpha\neq 1$. In the case of $\alpha>1$, its behaviour is known as superdiffusion transport due to the ballistic motion which could contribute to raising photovoltaic device efficiency \cite{guo2017long}. $\alpha<1$ characterizes the subdiffusion transport process due to the presence of exciton traps \cite{akselrod2014visualization,pandya2021exciton,burdett2010excited,delor2020imaging,yuan2017exciton}, which result from the energetic disorder, the nanoscale morphology, the domain boundary, defect scattering of material, or delocalization and symmetry of exciton. The nature is that the in-plane exciton transport is impeded, leading to the decrease of diffusivity. In general, subdiffusion is observed in biology and the other complex systems.

\subsection{the spin Hamiltonian}\label{subsec2.2}
In this part, the system studied will be simply illustrated, and the detailed state has been shown in previous paper \cite{hu2019improved}.

We take the solid rubrene as an archetype molecular crystal. The molecule system consists of two independent molecules. Excited by a photon, it can form a pair of excitons through SF process, i.e., a 4-electron spin system. For organic, the hyperfine interaction and the spin-orbit coupling are neglected. When an external magnetic field is applied, the total spin Hamiltonian of the system contains three parts,
\begin{eqnarray}
\hat{H}=\hat{H}_{{\rm Ze}}+\hat{H}_{{\rm zfs}}+\hat{H}_{{\rm ex}}. \label{Eq:total hamiltonian}
\end{eqnarray}
where $\hat{H}_{\rm Ze}$ describes the Zeeman splitting of the electron spin system in an external magnetic field $\textbf{B}$, $\hat{H}_{\rm zfs}$ is the total zero-field term from the spin-spin interaction of each electron-hole pair in two molecular, and $\hat{H}_{\rm ex}$ is the exchange interaction of the intermolecular spin-spin interaction. Then the spin Hamiltonian (\ref{Eq:total hamiltonian}) can be represented by the ordered basis \{$|xx\rangle$, $|xy\rangle,~\cdots,~|zz\rangle$\} of the triplet product states, where \{$|x\rangle$, $|y\rangle,~|z\rangle$\} correspond to the $X$, $Y$ and $Z$-axis of the our global coordinate frame defined, and $\textbf{B}=(0,0,B)$.

In order to display the effect of magnetic field on the fluorescence decay in SF, the overall singlet projection $|C^{l}_{s}|^{2}=|\langle S|\phi_{l}\rangle|^{2}$ need to be calculated below. 
$|S\rangle$ is the overall singlet state, and {$|\phi_{l}\rangle$} is the $l$-th eigenstate of the Hamiltonian (\ref{Eq:total hamiltonian}).

\subsection{the Kinetic Model with Time Dependent Coefficient on SF}\label{subsec2.3}
To investigate the effect of the AP on SF for low laser intensity conditions, the amended kinetic equations are introduced in this subsection based on the model shown by Fig. 3 of Ref. \cite{hu2019improved} except that the site number is set as 4 in molecule chain. The first site contains the ground singlet state ${\rm S_{0}}$, the overall singlet state ${\rm S_{1}}$, the associated triplet pair state $({\rm TT})_{l}$, and the spatially separated triplet pair state (T$\cdots$T)$_{l}$, while sites 2-4 contain a series of single triplets, T. In site 4, some of the triplet excitons are collected and cannot diffuse back. The equations that contain the parameters corresponding to the APs can be written as
\begin{subequations}\label{Eq:SF dynamics}
\begin{align}
 & \dot{N}_{{\rm S0}}=k_{{\rm rad}}\left(t\right)N_{{\rm S1}} \label{Eq:SF dynamics 1} \\
 & \dot{N}_{{\rm S1}}=-\left[k_{{\rm rad}}\left(t\right)+k_{{\rm SF}}\left(t\right)\right]N_{{\rm S1}}+k_{{\rm TF}}\left(t\right)\sum_{l=1}^{9}|C_{{\rm S}}^{l}|^{2}N_{{\rm (TT)}l}\\
 & \dot{N}_{{\rm (TT)}l}=k_{{\rm SF}}\left(t\right)|C_{{\rm S}}^{l}|^{2}N_{{\rm S1}}-\nonumber \\ &\left[k_{{\rm TF}}\left(t\right)|C_{{\rm S}}^{l}|^{2}+k_{{\rm dis}}\left(t\right)+k_{{\rm relax}}\left(t\right)\right]N_{{\rm (TT)}l}\nonumber \\
 & +\sum_{j\neq l}\frac{1}{8}k_{{\rm relax}}\left(t\right)N_{{\rm (TT)}j}+k_{{\rm comb}}\left(t\right)N_{{\rm (T...T)}l}\\
 & \dot{N}_{{\rm (T...T)}l}=k_{{\rm dis}}\left(t\right)N_{{\rm (TT)}l}-\nonumber \\ 
 &\left[k_{{\rm comb}}\left(t\right)+\frac{k_{{\rm hT}}\left(t\right)}{9}+k_{{\rm relax}}\left(t\right)\right]N_{{\rm (T...T)}l}\nonumber \\
 & +\sum_{j\neq l}\frac{1}{8}k_{{\rm relax}}\left(t\right)N_{{\rm (T...T)}j}+k_{{\rm h1}}\left(t\right)N_{{\rm T1}}\\
 & \dot{N}_{{\rm T1}}=\frac{k_{{\rm hT}}\left(t\right)}{9}\sum_{l=1}^{9}N_{{\rm (T...T)}l}-2k_{{\rm hl}}\left(t\right)N_{{\rm T1}}+k_{{\rm h2}}\left(t\right)N_{{\rm T2}}\\
 & \dot{N}_{{\rm T2}}=k_{{\rm h1}}\left(t\right)N_{{\rm T1}}-2k_{{\rm h2}}\left(t\right)N_{{\rm T2}}\\
 & \dot{N}_{{\rm T3}}=k_{{\rm h2}}\left(t\right)N_{{\rm T2}}
\end{align}
\end{subequations}

\begin{figure}
	\centering
	\includegraphics[width=8cm]{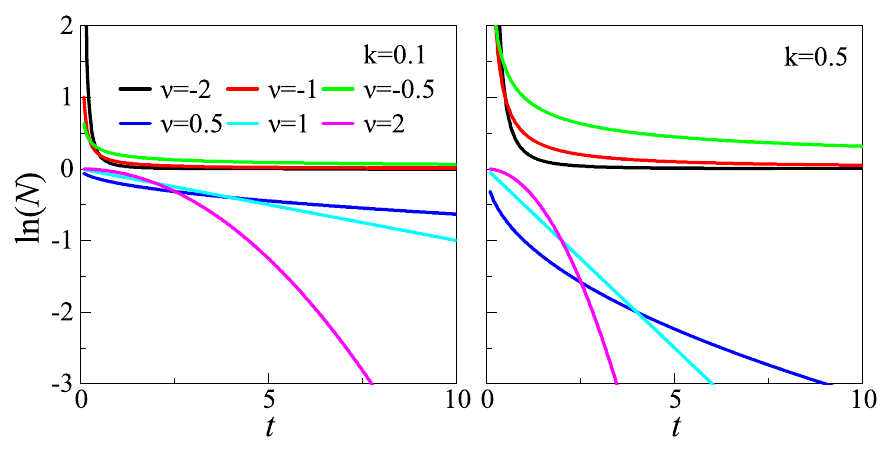}
	\caption{(color online) The time evolution of the solution of the Eq.(\ref{Eq:simplified differential}) for various values of parameters $k$ and $\nu$.}
	\label{FIG:simplified differential}
\end{figure}

\begin{figure*}[ht]
	\centering
	\includegraphics[width=16cm]{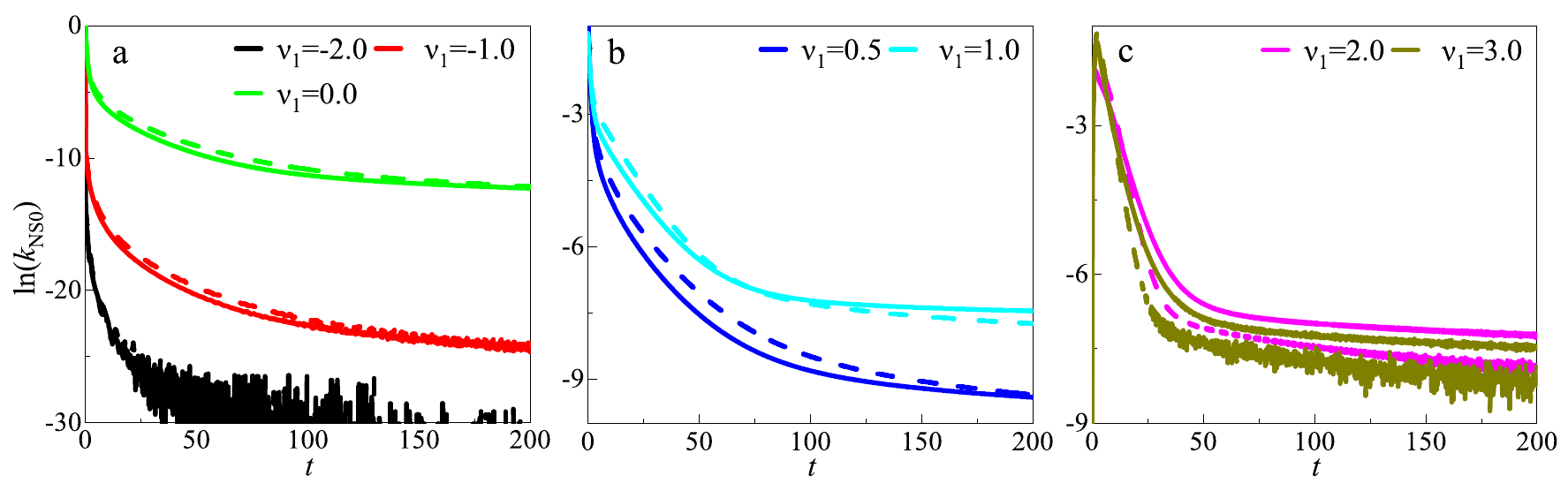}
	\caption{(color online) The FDs for different rate exponent $\nu_{1}$ within the singlet decay in the 200 ns time window, which are obtained by numerically solving Eqs.(\ref{Eq:SF dynamics}). The parameters used are illustrated in the text, unless otherwise noted below.}
	\label{FIG:niu1(4)}
\end{figure*}

\begin{figure*}
	\centering
	\includegraphics[width=16cm]{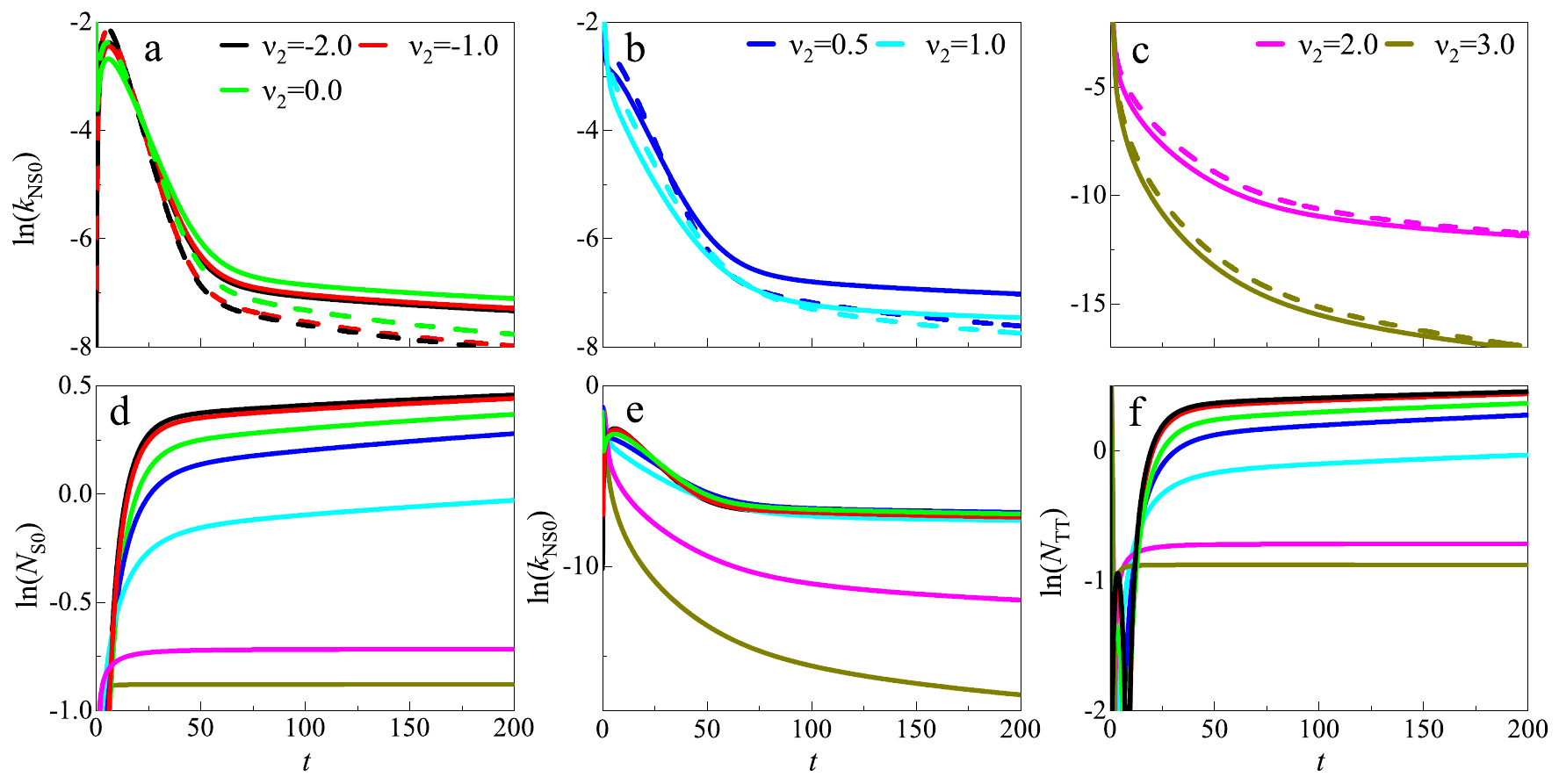}
	\caption{(color online) (a-c) The FDs for different rate exponent $\nu_{2}$ within the SF rate. (d-f) The time evolution of natural logarithm of the physical quantity $N_{\rm S0}$, $k_{\rm NS0}$, and $N_{\rm TT}$ for different $\nu_{2}$, respectively.}
	\label{FIG:niu2(5)}
\end{figure*}

\begin{figure*}
	\centering
	\includegraphics[width=16cm]{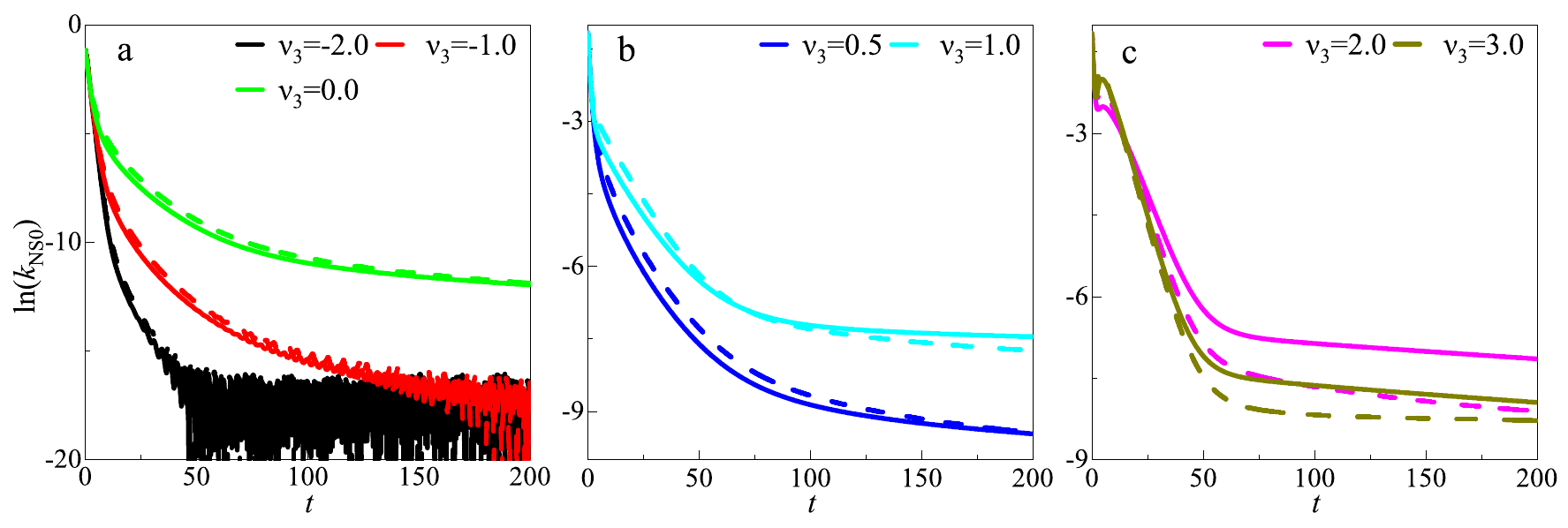}
	\caption{(color online) The FDs for different rate exponent $\nu_{3}$ within the triplet fusion rate.}
	\label{FIG:niu3(3)}
\end{figure*}

\begin{figure*}
	\centering
	\includegraphics[width=16cm]{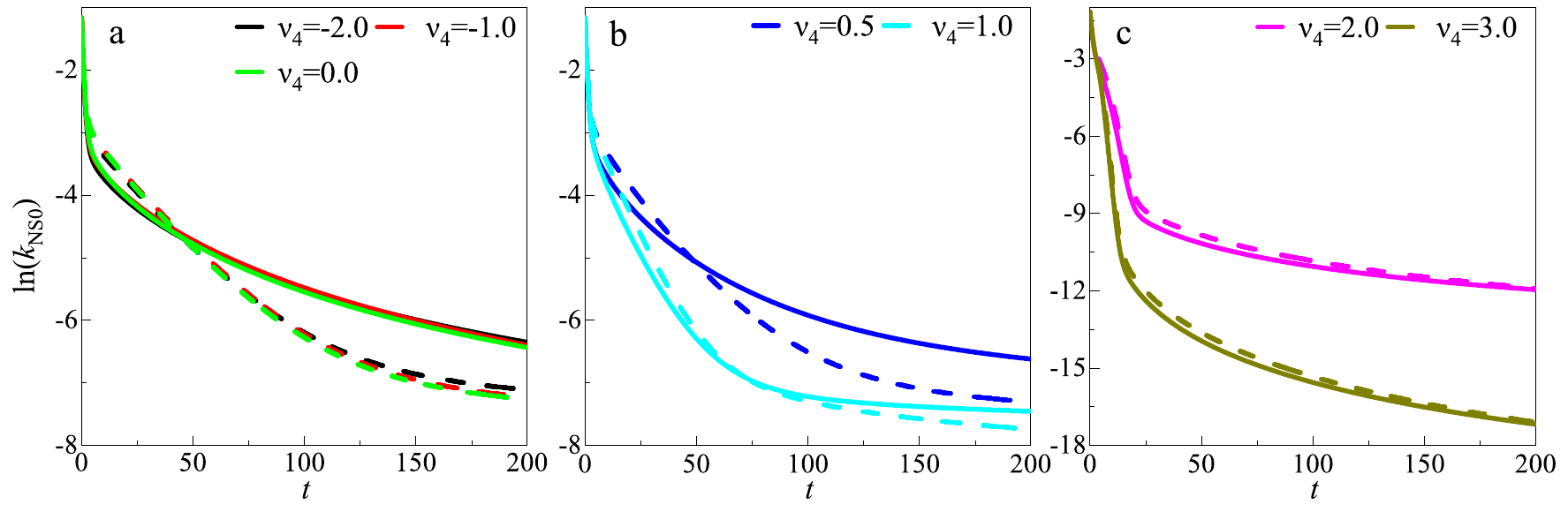}
	\caption{(color online) The FDs for different rate exponent $\nu_{4}$ within the dissociation rate of the associated triplet pair states.}
	\label{FIG:niu4(3)}
\end{figure*}

\begin{figure}
	\centering
	\includegraphics[width=8cm]{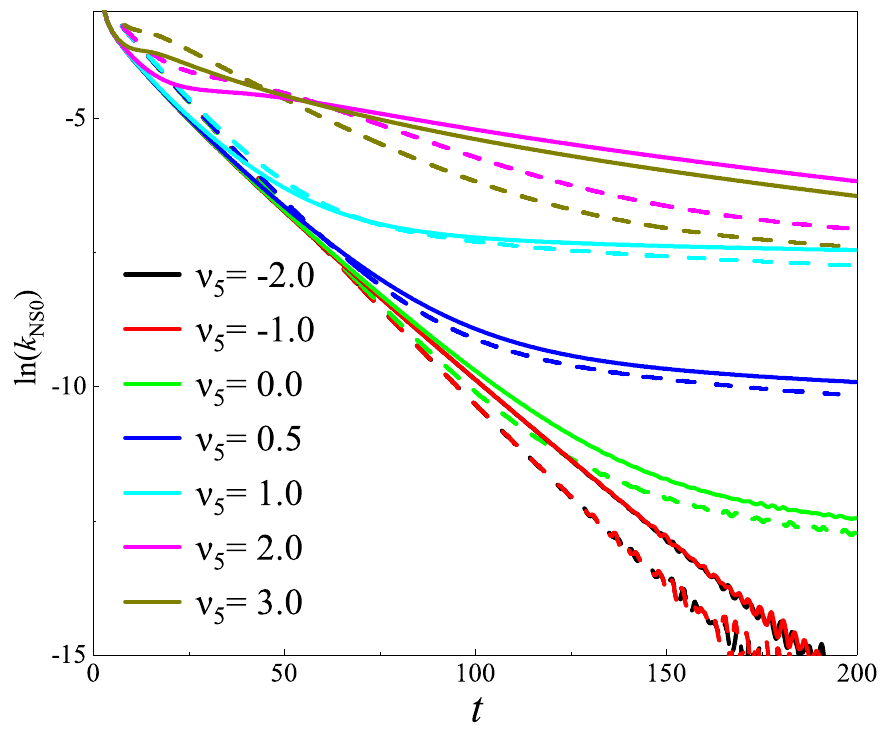}
	\caption{(color online) The FDs for different rate exponent $\nu_{5}$ within the combination rate of the dissociated triplet pair states.}
	\label{FIG:niu5(4)}
\end{figure}

\begin{figure*}
	\centering
	\includegraphics[width=16cm]{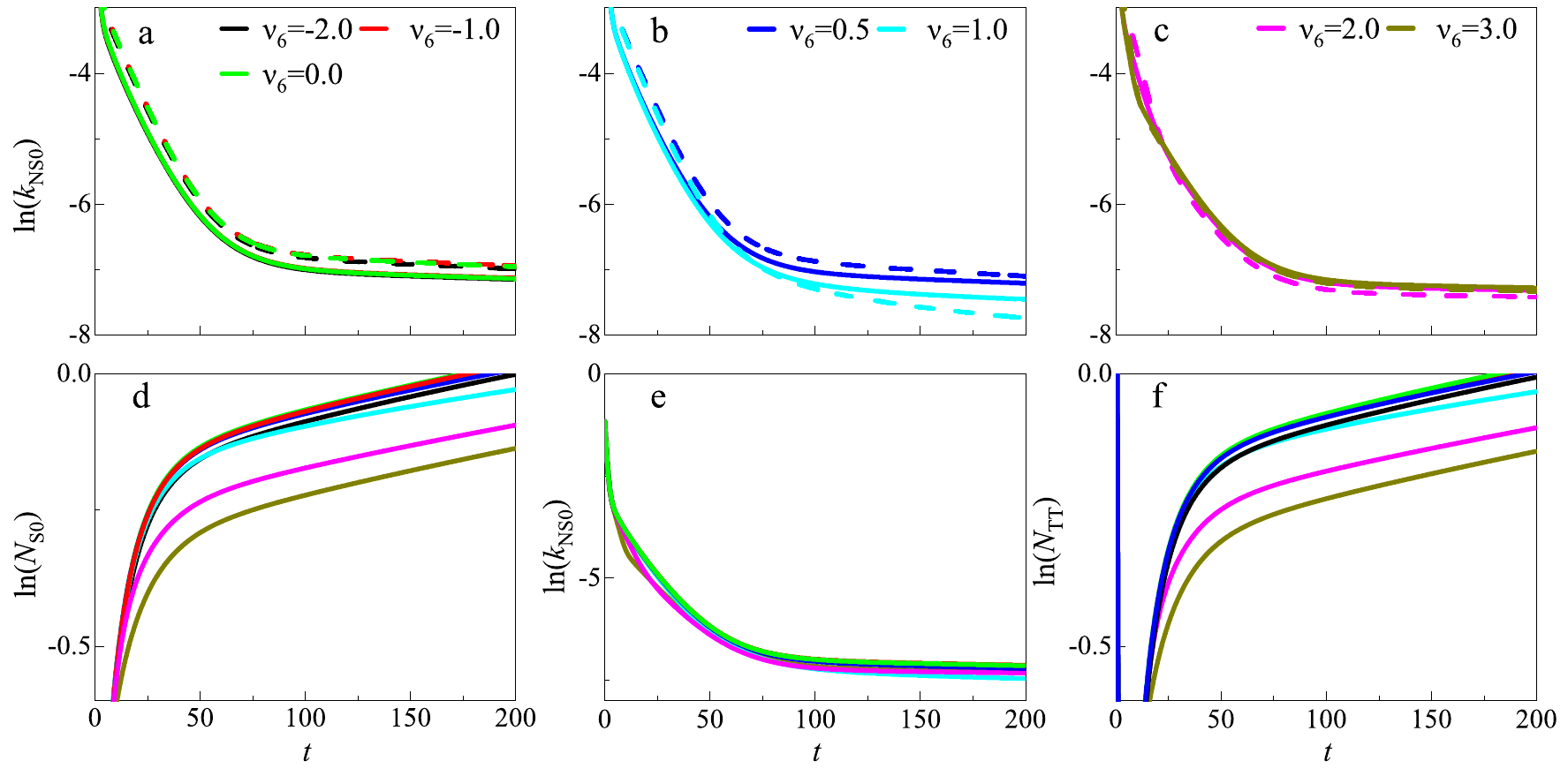}
	\caption{(color online) (a-c) The FDs for different rate exponent $\nu_{6}$ within the relaxation rate among triplet pair states. (d-f) The time evolution of natural logarithm of the physical quantity $N_{\rm S0}$, $k_{\rm NS0}$, and $N_{\rm TT}$ for different $\nu_{6}$, respectively.}
	\label{FIG:niu6(5)}
\end{figure*}

\begin{figure}
	\centering
	\includegraphics[width=8cm]{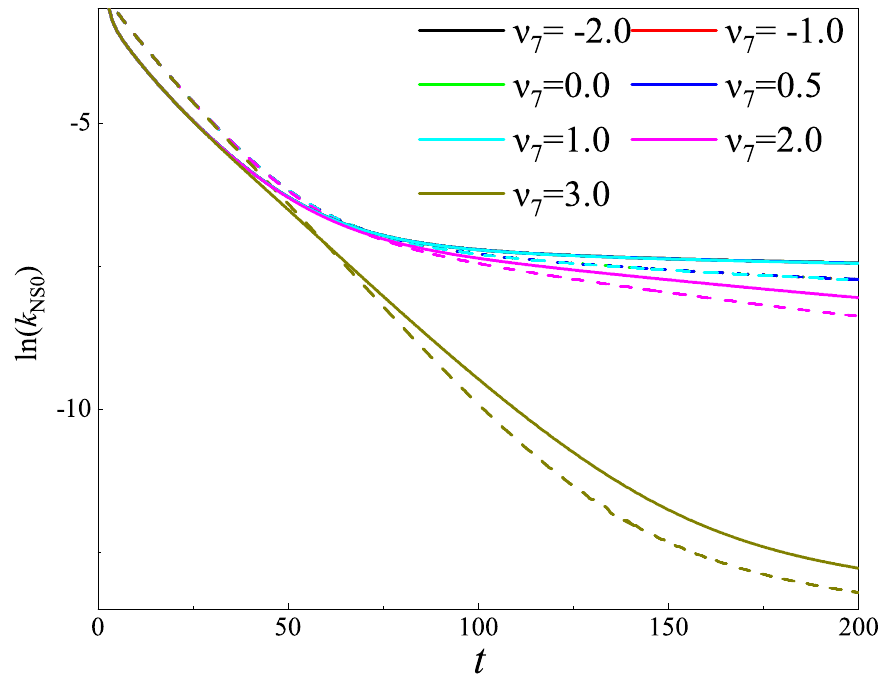}
	\caption{(color online) The FDs for different rate exponent $\nu_{7}$ within the diffusion rate due to the hopping of single triplet.}
	\label{FIG:niu7(4)}
\end{figure}

\begin{figure*}
	\centering
	\includegraphics[width=16cm]{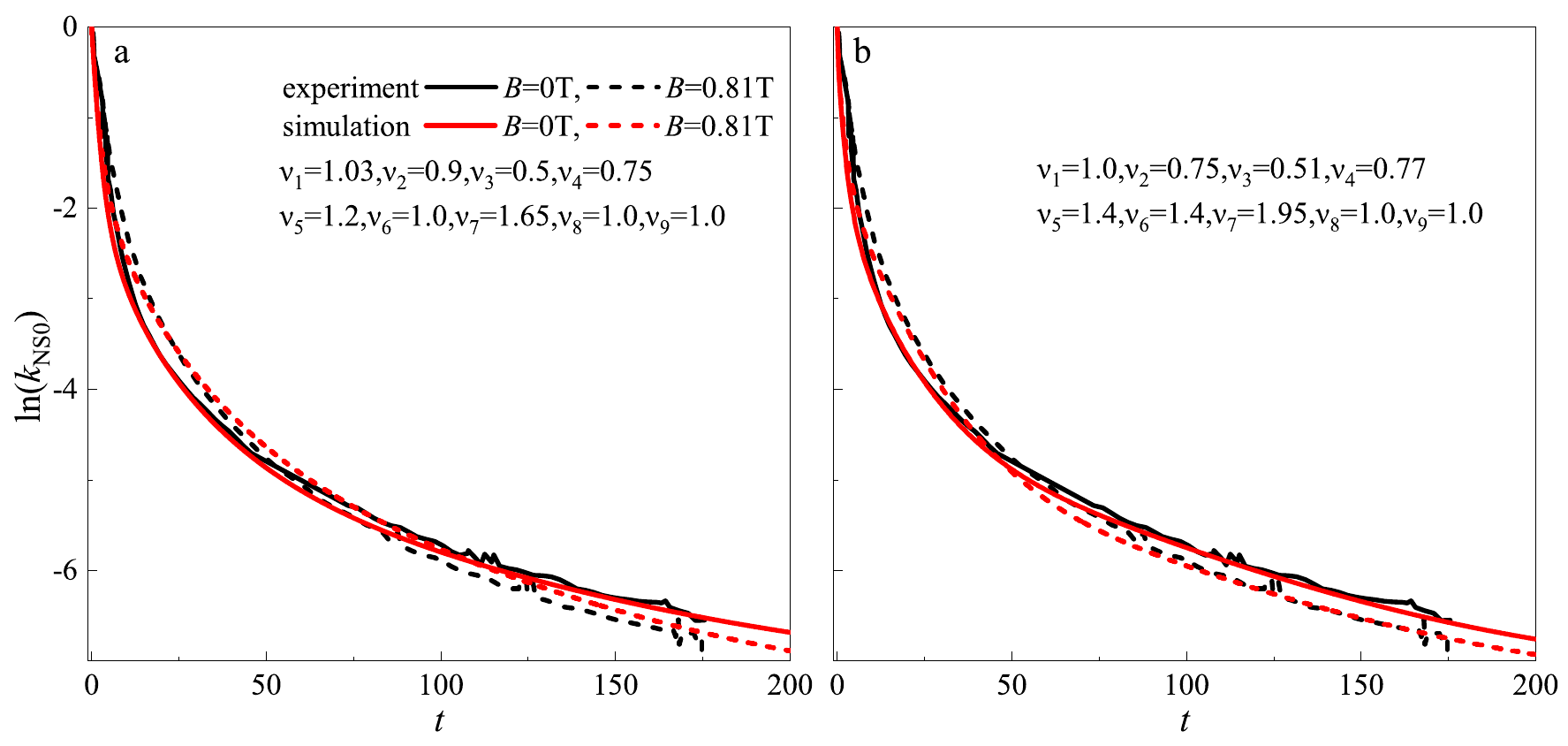}
	\caption{(color online) The optimal simulation on experimental results \cite{G.B.Piland.117.2013} for two groups of values of the parameters $\nu_{1-9}$ by Eqs.(\ref{Eq:SF dynamics}), the rate coefficients are $k_{1}=0.15$, $k_{2}=0.6$, $k_{3}=0.7$, $k_{4}=0.05$, $k_{5}=0.003$, $k_{6}=0.01$, $k_{7}=0.006$, $k_{8}=k_{9}=0.0005$. }
	\label{FIG:fitting11(2)_18}
\end{figure*}

\noindent where the dot denotes the derivative of population with respect to $t$. It is well known that the kinetic equations govern the time evolution of the population $N_{\rm S0}$, $N_{\rm S1}$, $N_{{\rm (TT)}l}$, $N_{{\rm (T...T)}l}$, and $N_{{\rm T1-T3}}$ of ${\rm S_{0}}$, ${\rm S_{1}}$, $({\rm TT})_{l}$, (T$\cdots$T)$_{l}$, and the single triplet state T at sites $2-4$, in which the dynamics coefficients with time dependence are $k_{{\rm rad}}\left(t\right)=k_{1}t^{\nu_{1}-1},\:k_{{\rm SF}}\left(t\right)=k_{{\rm 2}}t^{\nu_{2}-1},\:k_{{\rm TF}}\left(t\right)=k_{{\rm 3}}t^{\nu_{3}-1},\:k_{{\rm dis}}\left(t\right)=k_{4}t^{\nu_{4}-1},\:k_{{\rm comb}}\left(t\right)=k_{5}t^{\nu_{5}-1},\:k_{{\rm relax}}\left(t\right)=k_{6}t^{\nu_{6}-1}$ 
$k_{{\rm hT}}\left(t\right)=k_{7}t^{\nu_{7}-1},\:k_{{\rm hl}}\left(t\right)=k_{8}t^{\nu_{8}-1}$, and $k_{{\rm h2}}\left(t\right)=k_{9}t^{\nu_{9}-1}$,
$k_{i}$ is rate coefficient, and the parameter $\nu_{i}$ is called as the rate exponent by us in this paper, here $i=1,2,...9$. These dynamics coefficients $k_{{\rm rad}}\left(t\right)$, $k_{{\rm SF}}\left(t\right)$, $k_{{\rm TF}}\left(t\right)$, $k_{{\rm dis}}\left(t\right)$, $k_{{\rm comb}}\left(t\right)$, $k_{{\rm relax}}\left(t\right)$, $k_{{\rm hT,h1,h2}}\left(t\right)$ describe respectively the rate of radiative decay from the singlet state, the SF, the triplet fusion, the dissociation rate of the associated triplet pair states, combination rate of the dissociated triplet pair states, the transfer among the triplet pair states, and the diffusion rate of single triplet exciton due to the hopping among sites 1-4. Comparing the Eqs.(\ref{Eq:SF dynamics}) and the expressions of the dynamics coefficients with the Eq.(\ref{eq:diffusion}) and the Eq.(\ref{eq:diffusivity}) respectively, we can discover that the dynamics coefficient $k(t)$ and the rate exponent $\nu_{i}$ are respectively analogous to the diffusivity $D(t)$ and the diffusion exponent $\alpha$. Therefore, the values of $\nu_{i}$ also have three situations. Similarly, for $\nu_{i}=1$, the reaction in SF dynamics is normal process; for $\nu_{i}>1$, it is superprocess; for $\nu_{i}<1$, it is subprocess. For $\nu_{i}\neq1$, the $N_{\rm S1}$ and $N_{{\rm (TT)}l}$ follow an exponential evolution with non-one exponent. In mathematics, the relation between $k(t)$ and $t$ is power function when $\nu_{i}$ has a specific value. Yao Yao pointed out that the anomalous dynamics is caused by the nonlocal phonon, resulting in the nonlinear relationship
between the singlet and the triplet population \cite{yao2016coherent}, which has been verified
in an experiment \cite{zhang2014nonlinear}. According to this conclusion, we further suppose that the anomalous dynamics also can cause the nonlinear relationship among the others spin states. In addition to, it can be seen from the equations that the application of magnetic field changes the eigenstates of Hamiltonian (\ref{Eq:total hamiltonian}), leading to the change of the overall singlet projection $|C^{l}_{s}|^{2}$ of each eigenstate, hence affecting the solutions of Eqs.(\ref{Eq:SF dynamics}).

In Ref. \cite{hu2019improved}, the $\dot{N}_{\rm S0}$ is proportional to the $N_{\rm S1}$, and thus both $\dot{N}_{\rm S0}$ and $N_{\rm S1}$ can describe the fluorescence intensities of SF. However, according to Eq.(\ref{Eq:SF dynamics 1}) $\dot{N}_{\rm S0}(t)$ depends on both $k_{{\rm rad}}(t)$ and $N_{\rm S1}(t)$. Hence, in SF dynamics described by Eqs.(\ref{Eq:SF dynamics}) the evolution of the fluorescence intensities with $t$ can not be represented by $N_{\rm S1}(t)$ solely, but should be done by $\dot{N}_{\rm S0}(t)$. For convenience, $\dot{N}_{\rm S0}(t)$ is denoted by $k_{\rm NS0}$. The theoretical time evolution of the natural logarithm of time derivative of singlet state population ${\rm Ln}(k_{\rm NS0})$ represents the experimental radiation fluorescence intensity that comes from the transition ${\rm S}_{1}\rightarrow {\rm S}_{0}$. 
For the amorphous sample, the orientations of the magnetic axes of two independent rubrene molecules with respect to each other and an applied magnetic field \textbf{B} are arbitrary. Thus, we take the average value of fluorescence intensities for all orientations as precise value, which is called as total random situation. The following results are of total random, unless otherwise noted.

\section{The Results}\label{sec3}

\subsection{the Effects of Each AP on SF Dynamics}\label{subsec3.1}
Firstly, we investigate in detail the effects of each AP on SF dynamics by numerically solving the Eqs.(\ref{Eq:SF dynamics}) with the Hamiltonian (\ref{Eq:total hamiltonian}) in this subsection. The normalized initial condition is that all of the population are excited in the singlet ${\rm S_{1}}$ at the beginning, i.e., $N_{\rm S1}(0)=1$. When the rate exponent $\nu_{i}\neq1$, the $\nu_{1-6}$ denote respectively the APs of the radiative decay from the singlet state, the SF, the triplet fusion, the dissociation of the associated triplet pair states, combination of the dissociated triplet pair states, and the relaxation among the triplet pair states in site 1. Besides, the $\nu_{7-9}\neq 1$ denote the anomalous diffusion of single triplet exciton due to the hopping motion between neighbouring sites in sites 1-4. Particularly, the dynamics of $\nu_{i}=1$ is normal process. In fact, our aim is to investigate the effects of the APs on SF dynamics. For the situation of total random, the average values of the ground singlet state population $N_{\rm S0}$ and the singlet state population $N_{\rm S1}$ are obtained by selecting a spherically uniform distribution of the orientations of two molecules and the magnetic field. The values of the related parameters must be determined for theoretical simulation. The ways that obtain dynamics coefficients $k(t)$ are simply summarized in subsection 3.2 of Ref. \cite{hu2019improved}. In addition to, the rate exponent $\nu_{i}$ can be obtained by fitting to the dynamics that simulates by full-quantum theory \cite{yao2016coherent} and to experimental results on fluorescence decay. The results investigated on the effects of $\nu_{i}$  can well help ourselves perform the theoretically fitting on experimental result. Further, it can also provide a theory foundation for optimizing and designing photovoltaic devices. In the following simulation, the parameters used are set as the magnetic field $B=0.81~T$, the product $g\beta=500~{\rm m^{-1}T^{-1}}$ of the magnetic field coupling constant $\beta$ and the gyromagnetic ratio $g$, the parameters $D=-0.62~{\rm m^{-1}}$ and $E=2.48~{\rm m^{-1}}$ of zero-field splitting, and the coupling strength of the exchange interaction  $X=0.01~{\rm m^{-1}}$ in total spin Hamiltonian (\ref{Eq:total hamiltonian}). Besides, the relevant rate coefficients are the values of the rate constants got by optimal fitting in Ref. \cite{hu2019improved}, that is $k_{1}=0.15~{\rm ns}^{-1},~k_{2}=0.8~{\rm ns}^{-1},~k_{3}=0.7~{\rm ns}^{-1},~k_{4}=0.05~{\rm ns}^{-1},~k_{5}=0.003~{\rm ns}^{-1},~k_{6}=0.01~{\rm ns}^{-1},~k_{7}=0.0005~{\rm ns}^{-1},~k_{8}=0.0005~{\rm ns}^{-1},~k_{9}=0.0005~{\rm ns}^{-1}$. The relevant results are shown in Fig. \ref{FIG:niu1(4)}-\ref{FIG:niu7(4)}.
 
In order to analysing the SF dynamics described by the Eqs.(\ref{Eq:SF dynamics}), we give a simplified non-couple differential equation as 
\begin{equation}
\dot{N}=-kt^{\nu-1}N \label{Eq:simplified differential}
\end{equation}
As shown in Fig. \ref{FIG:simplified differential}, the time evolution of the solution $N(t)$ of the Eq.(\ref{Eq:simplified differential}) is linear, and is a normal exponential decay for $\nu=1$ and $k>0$. However, the non-linear evolution is sub-decay process for $\nu <1$, i.e., the decay is decelerated. On the other hand, it is super-decay process for $\nu>1$, i.e., the decay is accelerated. Specially, for $\nu=0$ the form of $N(t)$ is different from that for $\nu\neq 0$, but the solution of Eqs.(\ref{Eq:SF dynamics}) is common because of the coupling of many equations and is no longer illustrated below. It also can be seen that the influence of $\nu$ on $N(t)$ is enlarged with $k$ increasing. Besides, the increase of $k$ weakens the decay for $\nu<0$ and strengthen that for $\nu>0$. Although the evolution of the solution of the Eqs.(\ref{Eq:SF dynamics}) is composite exponential evolution due to the coupling among these equations, the foundation to analyse it is the characters of $N(t)$.

It is well-known that singlet excited has two approaches to decay, one is singlet exciton decay to ground singlet $S_{1}\rightarrow S_{0}$ whose rate is determined by parameters $k_{1}$ and $\nu_{1}$, and the other is SF process $S_{1}\rightarrow (TT)_{l}$ whose rate is determined by $k_{2}$ and $\nu_{2}$. The two processes are competitive each other, and the effect of its rate exponent $\nu_{1,2}$ on whole SF dynamics are shown in Fig. \ref{FIG:niu1(4)} and \ref{FIG:niu2(5)}. It is obvious in Fig. \ref{FIG:niu1(4)} that $\nu_{1}$ is so small and large for $\nu_{1}=-2$ and 3 respectively that the evolution of the system is unstable. Therefore, in SF the obstructing or excessively fast local process can break the continuousness of system. Of course, a relatively weak instability are shown for $\nu_{1}=-1$ and 2. We also can find that the saturation value that the singlet decay arrives increases for $\nu_{1}<1$ and decrease for $\nu_{1}>1$ with $\nu_{1}$ increasing. In addition to, the increase of $\nu_{1}$ changes the effect of magnetic field on SF dynamics, namely the cross point of FD line with and without magnetic field shifts forward. The strengthening and weakening of prompt and delayed fluorescence are main performance of magnetic field effect on SF. In FD lines the character is that the cross point emerges early or late. We will primarily concern the cross point about magnetic field effect below. It can be seen in Fig. \ref{FIG:niu2(5)}(b) that the effect of $\nu_{2}$ including on the saturation value and the cross point is opposite to that of $\nu_{1}$ for $0<\nu_{2}<1$ due to the competition between SF and singlet decay. In Fig. \ref{FIG:niu2(5)}(c) the effects of $\nu_{1}$ on SF dynamics are qualitatively consistent with that of $\nu_{2}$ for $\nu_{2}>1$, which is beyond expectation and breaks the routine due to super-SF process. Particularly, the fluorescence intensities increase at first and then decrease for $\nu_{2}<0$ as shown in Fig. \ref{FIG:niu2(5)}(a). In order to explain this, the time evolutions of natural logarithm of the physical quantity $N_{\rm S0}$, $N_{\rm TT}$, and $k_{\rm NS0}$ are displayed for zero magnetic field $B=0$ in Fig. \ref{FIG:niu2(5)}(d-f). It can be seen at the time when the peaks of ${\rm Ln}(k_{\rm NS0})$ emerge in Fig. \ref{FIG:niu2(5)}(e) that the evolutions of ${\rm Ln}(N_{\rm TT})$ also present peaks in Fig. \ref{FIG:niu2(5)}(f). It is easy to be understood dynamically that when $\nu_{2}<0$, the decay of $N_{\rm S1}$ is decelerated, leading to that the population of $S_{1}$ collect, intensifying the processes of $S_{1}\rightarrow S_{0}$ and $S_{1}\rightarrow (TT)_{l}$, and thus the slopes of ${\rm Ln}(k_{\rm NS0})$ increase temporarily as shown in Fig. \ref{FIG:niu2(5)}(a). On the other hand, the SF is the inverse process of triplet fusion, and thus the effect of $\nu_{1}$ on SF dynamics is qualitatively consistent with that of $\nu_{3}$ as shown in Fig. \ref{FIG:niu3(3)}.

Fig. \ref{FIG:niu4(3)} and \ref{FIG:niu5(4)} display the influence of rate exponent $\nu_{4}$ and $\nu_{5}$ within the dissociation and combination process of triplet pair state on SF dynamics respectively. The two processes are mutually inverse, which is similar to the relation between SF and triplet fusion, as the consequence the influences of $\nu_{4}$ and $\nu_{5}$ also are inverse.  The changes of both $\nu_{4}$ and $\nu_{5}$ influence the delayed fluorescence, but there are a little difference. The influences for $\nu_{4}>1$ and for $0\leq\nu_{5}\leq 1$ are significant. Quantitatively, according to Eq. (\ref{Eq:simplified differential}) the no difference for $\nu_{4}\leqslant0$ attributes to too small rate coefficient $k_{4}$ in Fig. \ref{FIG:niu4(3)}. We also can see the same situations in Fig. \ref{FIG:niu5(4)}-\ref{FIG:niu7(4)} because of the same reason. Therefore, if one need to adjust FD line shape for $\nu_{4-7}\leqslant0$, one of the effective ways is to turn $k_{4-7}$ large. As we known that the dissociation and combination processes of triplet pair state can respectively prompt and restrain SF process, and thus the influences of $\nu_{4}$ and $\nu_{5}$ shown in Fig. \ref{FIG:niu4(3)} and \ref{FIG:niu5(4)} are logical.

The effects of rate exponent $\nu_{6}$ of the relaxation process of triplet pair state and $\nu_{7}$ of the diffusion process of single triplet exciton due to the hopping between sites 1 and 2 are shown in Fig. \ref{FIG:niu6(5)} and \ref{FIG:niu7(4)}. As shown in Fig. \ref{FIG:niu6(5)}(b), it is obvious just for $0<\nu_{6}<1$. For illustrating it, we present the evolutions of ${\rm Ln}(N_{\rm S0})$, ${\rm Ln}(N_{\rm TT})$, and ${\rm Ln}(k_{\rm NS0})$ for $B=0$. Although the influence of $\nu_{6}$ on FD is faint, those of $\nu_{6}$ on the population of singlet $N_{\rm S0}$ and triplet pair state $N_{\rm TT}$ are significant. In other word, $\nu_{6}$ mainly affects the population being saturation of $S_{0}$ and $(TT)_{l}$, but not the rate of radiative decay of singlet. In the design of photocells, the diffusion of exciton is a important factor to affect efficiency. However, in SF the influence of rate exponent $\nu_{7}$ of the diffusion process of single triplet exciton is obvious just for super-diffusion $\nu_{7}>1$, and mainly behaves at delayed fluorescence as shown in Fig. \ref{FIG:niu7(4)}. The saturation values are lifted with the decrease of $\nu_{7}$ for $\nu_{7}>1$. According to this results, large diffusivity is beneficial to the SF process, on the other hand to the collection of triplet exciton. Because the SF dynamics does not behave the influences of $\nu_{8}$ and $\nu_{9}$, the results is not presented in figure. Therefore, the diffusion beyond the site SF occur has no influence on SF dynamics, which perhaps is attributed to that our model is imperfect. 

\subsection{the Optimal Fitting on Experiment}\label{subsec3.2}
Based on the results analysed in SubSec. \ref{subsec3.1}, in this subsection we give the optimal fitting of Eqs. (\ref{Eq:SF dynamics}) on experimental data about time-resolved FD of amorphous rubrene thin films applied an external strong magnetic field in Ref. \cite{G.B.Piland.117.2013}. The core of fitting is that the optimal rate coefficients $k_{i}$ and exponents $\nu_{i}$ are manually found out, in which the values of the other parameters are same as that in SubSec. \ref{subsec3.1}. The FDs solved by Eqs. (\ref{Eq:SF dynamics}) with these values of $k_{i}$ and $\nu_{i}$ are the closest to the experimental that. The procedure finding $k_{i}$ and $\nu_{i}$ is similar to that finding rate constant $k$ in Ref. \cite{hu2019improved}, here specially not repeat it. In order to simulate conveniently experimental results, the experimental results are shown in Fig. \ref{FIG:fitting11(2)_18} through software, whose initial points are shifted to $t=0$. Fig. \ref{FIG:fitting11(2)_18} presents two groups $\nu_{i}$ attached to optimal fitting whose line shape [Fig. \ref{FIG:fitting11(2)_18}(a)] and cross point [Fig. \ref{FIG:fitting11(2)_18}(a)] are nearly consistent with the experimental results respectively. For more accurate to illustrate the fitting, the root mean square deviations of the two FDs simulated relative to experimental data are calculated as $\sigma_{1}=0.23$, $\sigma^{\prime}_{1}=0.21$ and $\sigma_{2}=0.25$, $\sigma^{\prime}_{2}=0.22$, where the marks without and with prime denote the situations of zero and strong magnetic field respectively. Nevertheless, its values are $\sigma=0.36$, $\sigma^{\prime}=0.3$ fitted by the model without considering AP (i.e., $\nu_{i}=1$) in Ref. \cite{hu2019improved}. Obviously, the FDs of optimal fitting through Eqs. (\ref{Eq:SF dynamics}) are more closer to experimental results than that through equations without AP. According to the consistence and the two groups of values fitted of $\nu_{i}$, one can conclude that the APs exist in real SF dynamics. In the first group (Fig. \ref{FIG:fitting11(2)_18}(a)), the APs contain the subprocesses of the SF ($\nu_{2}=0.9$), the triplet fusion ($\nu_{3}=0.5$), the dissociation of the associated triplet pair states ($\nu_{4}=0.75$), and the superprocesses of the radiative decay of singlet ($\nu_{1}=1.03$), the combination of the dissociated triplet pair states ($\nu_{5}=1.2$), the diffusion between sites 1 and 2 ($\nu_{7}=1.65$). The others also is noticeable. Therefore, the theory model with AP provides more precise foundation for experiment and reveals more comprehensive reaction mechanism.

\section{Conclusions}\label{sec4}
In conclusion, the singlet formed by a pair exciton excited by photon can undergo SF process, generating triplet pair and further single triplet exciton, which generally is described by a kinetic equations governing each spin state of 4-spin system. We introduce APs in a series of processes relating SF, including the radiative decay from the singlet state, the SF, the triplet fusion, the dissociation of the associated triplet pair states, the combination of the dissociated triplet pair states, the relaxation among the triplet pair states, and the diffusion of single triplet exciton due to the hopping motion between neighbouring sites in molecule chain, which are described by the time dependent coefficient with non-normal exponential evolution in the kinetic equations. Based on the kinetic equations, the effects of the APs of these processes on SF dynamics are investigated. The results show that these effects are various, which are beneficial to adjust SF dynamics. Therefore, it provide a theoretical foundation for experimentally designing photoelectric conversion device with high efficiency. Besides, according to the results, the optimal simulations are performed on experimental data about time-resolved FD of amorphous rubrene thin films applied an external strong magnetic field in Ref. \cite{G.B.Piland.117.2013} by the mended kinetic equations. The consistence between the experimental results and the simulation with AP exceeds previous that with normal processes of SF dynamics in Ref. \cite{hu2019improved}. The result suggests that the APs exist in SF dynamics, and reveals more comprehensive reaction mechanism in SF, including dynamical instability, subprocess and superprocess. The kinetic equations are helpful to understand SF process and the factor affecting efficiency of photoelectric conversion device. In consequence, we expect that it could provide some insights for its application on solar energy harvesting. Of course, there are some defects in the model, such as the influence of diffusion of position beyond SF occur on SF can not present. It should be hopeful to overcome the insufficient through full quantum theory.

\section{Acknowledgements}\label{sec5}
This work is supported by the Academic Ability Promotion Foundation for Young Scholars of Northwest Normal University in China under the Grant No. NWNU-LKQN2019-19, Regional Science Foundation of China under Grant No. 12164042, National Natural Science Foundation of China under Grant No. 12104374, and Natural Science Foundation of Gansu Province under Grant No. 20JR5RA526.



\end{document}